\documentclass[prc,aps,nofootinbib,showkeys,showpacs,twocolumn]{revtex4}
\usepackage{epsfig}
\usepackage{graphicx}
\usepackage{amssymb}
\usepackage{color}
\usepackage{amsmath}

\DeclareMathOperator{\Erf}{Erf}

\begin{document}

\title{Semi-contact 3-body interaction for nuclear density functional theory }
  
\author{Denis Lacroix} \email{lacroix@ipno.in2p3.fr}
\affiliation{Institut de Physique Nucl\'eaire, IN2P3-CNRS, Universit\'e Paris-Sud, F-91406 Orsay Cedex, France}
\author{K. Bennaceur}\email{bennaceur@ipnl.in2p3.fr}
\affiliation{Universit\'e de Lyon, F-69003 Lyon, France; Institut de Physique Nucl\'eaire de Lyon,
CNRS/IN2P3, Universit\'e Lyon 1, F-69622 Villeurbanne Cedex, France}
\affiliation{Department of Physics, PO Box 35 (YFL), FI-40014 University of Jyv\"askyl\"a,
Finland}
\date{\today}
\begin{abstract}
To solve difficulties related to the use of  nuclear density functional theory applied in its beyond mean-field
version, we introduce a semi-contact 3-body effective interaction. We show that this interaction is a good candidate to
replace the widely used density dependent effective interaction. The resulting new functionals are able to
describe symmetric, neutron, polarized and neutron polarized nuclear matter as well as the effective mass properties
simultaneously.
\end{abstract}

  

\maketitle

Strongly interacting fermions can exhibit peculiar behavior in the infinite
system limit from very dilute to very dense matter. In such systems, the zero
temperature equation of state (EOS), or energy per particle $E/A$, can be displayed in terms of the densities of
various constituents. Rather commonly, specific medium effects
take place and lead to an energy that is a functional that
depends on non-integer powers of the density.
This stems from the fact that a natural expansion parameter for homogenous infinite system is the
Fermi momentum $k_F$  that is proportional to $\rho^{1/3}$, where 
$\rho$ is the matter density.
A seminal example is the Lee-Yang formula for $E/A$
that holds for any diluted Fermi systems with short range
interaction~\cite{Lee57,Fet03} and contains a term proportional to $\rho^{4/3}$.
A more recent example is the universal class of Fermi systems at unitarity,
where a $\rho^{2/3}$ dependence~\cite{Bul07} is supported by ab-initio
calculations.
Cold atoms or neutron matter at low density can enter into this class of
systems~\cite{Gez08,Gez10}. In phenomenological approaches, as in the nuclear Density Functional Theory (DFT)~\cite{Ben03}, this density dependence of the functional has to be mimicked
one way or the other. Indeed, in nuclear systems, an indirect evidence
of the necessity to have a term that behaves like $\rho^{1+\alpha}$ with
$1/6 \leqslant \alpha \leqslant 1/3$ in the energy, is the difficulty to reproduce both the
compression modulus of nuclear matter and the quasiparticles effective
mass at saturation with integer powers of the density only~\cite{Ben03}. 

A clear view of the density dependence of the energy is crucial to design
an accurate DFT for the considered many-body problem. 
In finite systems, it might be necessary
to go beyond the mean-field approximation by accounting for quantum fluctuations in
collective space and restoring some symmetries that were broken to include correlations into a compact functional.
An example is the treatment of superfluidity in small superconductors or nuclei that can be included by 
breaking the U(1) symmetry~\cite{Rin80,Bla86}. 
It was shown recently  that symmetry restoration within
DFT requires a functional that is strictly derived from a N-body Hamiltonian
to avoid pathologies in the energy of the symmetry restored states
(see for instance~\cite{Lac09,Ben09,Dob07d}). It is for instance not possible to use a functional 
that depends on non-integer powers of the
density~\cite{Dug09}, due to the necessity to extend the functional theory to the complex plane.
The afrementioned pathologies are rooted in the violation of the Pauli principle and the occurrence of self-interaction.
This has renewed the interest in deriving a DFT from a
pseudo-potential approach. In that case, the Hartree-Fock expression obtained
from an
effective Hamiltonian is used to provide the functional whose parameters have
to be fitted -- at the mean-field level or beyond -- on data.

Recently, efforts have been made 
to extend functionals based on zero-range interactions of increasing complexity 
to overcome the difficulties arising from beyond mean-field calculations~\cite{Was12,Sad13}.
While very useful to get a Local Density Approximation (LDA) for the
functional, it is not yet clear if contact interactions can be used to provide a
convenient solution to the problem. For this reason, DFT based on finite range
pseudo-potential might appear as an alternative solution~\cite{Dob12a,Ben13,Rai14}.
A first step in that
direction has been made in Ref.~\cite{Gez10b} where a 3-body force was proposed 
to describe fermions in the low density limit.

Here, we explore the possibility to use a
density-independent 3-body interaction to get a functional that mimic non-integer power dependences. By imposing
the translational invariance, a 3-body 
interaction can only depend on the Jacobi coordinates ${\mathbf r_{ij}} = ({\mathbf r}_{i} - {\mathbf r}_{j})$, ${\bf R}^k_{ij} = {\mathbf r}_{k} - 
({\mathbf r}_{i} + {\mathbf r}_{j})/2$ where $({\mathbf r}_i, {\mathbf r}_j, {\mathbf r}_k)$ are
the coordinates of three particles, $(i,j,k)$ in an arbitrary frame. A general 3-body interaction can a priori have a finite range in both these coordinates. 
Below, we consider the case where the interaction is a contact interaction in ${\bf R}^k_{ij}$ and call it a 3-body semi-contact interaction.
It has the advantage
to be more flexible than a 3-body contact interaction.
To characterize it further, we consider three particles
with spin (and possibly isospin) degrees of freedom
that interact through $\bar v_{ijk} = (v_{ijk}+v_{ikj} +v_{ki j})/3$ with
\begin{eqnarray}
v_{ijk} &=& \Big\{ V_{0}(r) + V_\sigma ( r )
P_\sigma + V_\tau (r) P_\tau + V_{\sigma \tau} ( r ) P_\sigma P_\tau \Big\} \nonumber \\
& \times &\delta \left({\mathbf r}_k -\left[ \frac{{\mathbf r}_i + {\mathbf r}_j}{2}\right] \right), \label{eq:3bodycontact}
\end{eqnarray} 
where the short-hand notation $r=|{\mathbf r}_{i} - {\mathbf r}_{j}|$ is used
while $P_\sigma$ (resp. $P_\tau$) exchanges the projections of spin (resp. isospin)
of the particles $i$ and $j$. Note that, the interaction can be used without
isospin exchange operators in case it is applied to fermions with spin
degrees of freedom only by omitting the two terms with $P_\tau$ in Eq.~(\ref{eq:3bodycontact}). The functions $V_\alpha( r)$ with $\alpha=0$, $\sigma$,
$\tau$ and $\sigma\tau$ correspond to the 2-body interaction acting in
different channels. For simplicity, we assume that they can be written
in terms of a smooth normalized function $g$ as $V_\alpha ( r) = v_\alpha g( r)$
where $v_0$, $v_\sigma$, $v_\tau$ and $v_{\sigma\tau}$ are independent strength
parameters.
\begin{table}[htbp]
  \centering 
  \begin{tabular}{| c || c c | c c | c c | c c |}
\hline
& $c_1^{\rm SM}$ & $c_2^{\rm SM}$   & $c_1^{\rm NM}$ &  $c_2^{\rm NM}$ & $c_1^{\rm PM}$ &  $c_2^{\rm PM}$ 
& $c_1^{\rm PNM}$ &  $c_2^{\rm PNM}$  \\
\hline\hline
 $v_0$              & 1    & -1/4 &    1 &  -1/2  & 1 &  -1/2 & 1 &  -1 \\
 $v_\sigma$             & 1/2 & -1/2 &  +1/2 & -1  &  1 &  -1/2  & 1&  -1 \\
 $v_\tau$               & 1/2 & -1/2 &     1    & -1/2 &  1/2 & -1  & 1&  -1\\
$v_{\sigma \tau}$    & 1/4 &   -1  &  +1/2 & -1 &  1/2 &  -1 & 1&  -1 \\
\hline
\end{tabular}
  \caption{Correspondence between the coefficients $(c_1 ,c_2)$ and the interaction strengths. Table gives coefficient $y_{k\alpha}$ defined 
  as $c_k = \sum_k y_{k\alpha} v_\alpha$ are reported here.}
  \label{table:vb}
\end{table}

The functional associated to the semi-contact 3-body interaction is directly obtained from its Hartree-Fock expression for the energy. 
We focus here on infinite systems and introduce the 
Fermi momentum $k_F$ and the density of the Fermi gas $\rho = d\,k^3_F /(6\pi^2)$
where $d$ is the degeneracy 
that depends on the specific situation. We consider the four following cases: symmetric nuclear matter (SM) with $d=4$,
neutron matter (NM) and spin polarized matter (PM) with $d=2$ and spin
polarized neutron matter (PNM) with $d=1$. A lengthy but straightforward
calculation shows that the energy per particle can be written
\begin{eqnarray}
&&\frac{{E}^{\rho\rho\rho}}{A} =  c_1 \frac{\rho^2}{6} \!\int\! \mathrm d^3r \,g( r ) \left[ 1 - c_3  f(k_F r /2)^2 \right] \label{eq:edf}  \\
&&+c_2 \frac{\rho^2}{6}  \!\int\! \mathrm d^3r \,g(r) \left[  f (k_F r)^2 - c_3  f (k_F r) f(k_F r /2)^2 \right]\,,  \nonumber
\end{eqnarray} 
where the function $f$ is expressed in terms of the first spherical Bessel
function as $f(x) = 3 j_1(x)/x$.  The coefficients $c_1$, $c_2$ and $c_3$ depend
on the specific channel. One has $c_3 = 1/2$ for SM, $1$ for both NM and PM and
$2$ for PNM.  
The correspondences between the coefficients $c_1$ and $c_2$ and the 
parameters $v_\alpha$ are given in table~\ref{table:vb}. Eq.~(\ref{eq:edf})
can serve, given a specific 
function $g$ associated to the range $a$, to get expansions in powers of $(k_F a)$  and obtain the low density behavior. Note that, 
if a 2-body interaction is used with same finite range part as in Eq.~(\ref{eq:3bodycontact}), the energy is identical to Eq.~(\ref{eq:edf}) with $c_3=0$ and $\rho^2/6 \rightarrow \rho/2$. In this case, the functional will be denoted by ${E}^{\rho\rho}$. In SM,
the limit for a zero-range 2-body and 3-body interaction is obtained by setting $ g( r ) = \delta ( {\mathbf r})$ and gives
  \begin{eqnarray}
\frac{{E}^{\rho\rho}_{\rm zero}}{A} = \frac{3}{8} t_3 \rho, ~~
\frac{{E}^{\rho\rho\rho}_{\rm zero}}{A} =\frac{t_3}{8} (1-c_3) \rho^2, \label{eq:3zero}
\end{eqnarray}
with $t_3 = 4 (c_1 +c_2)/3$.  Note that for NM, PM and PNM, ${E}^{\rho\rho\rho}_{\rm zero}$ cancels out as expected. 

\begin{figure}[htbp]
\includegraphics[width=0.7\linewidth]{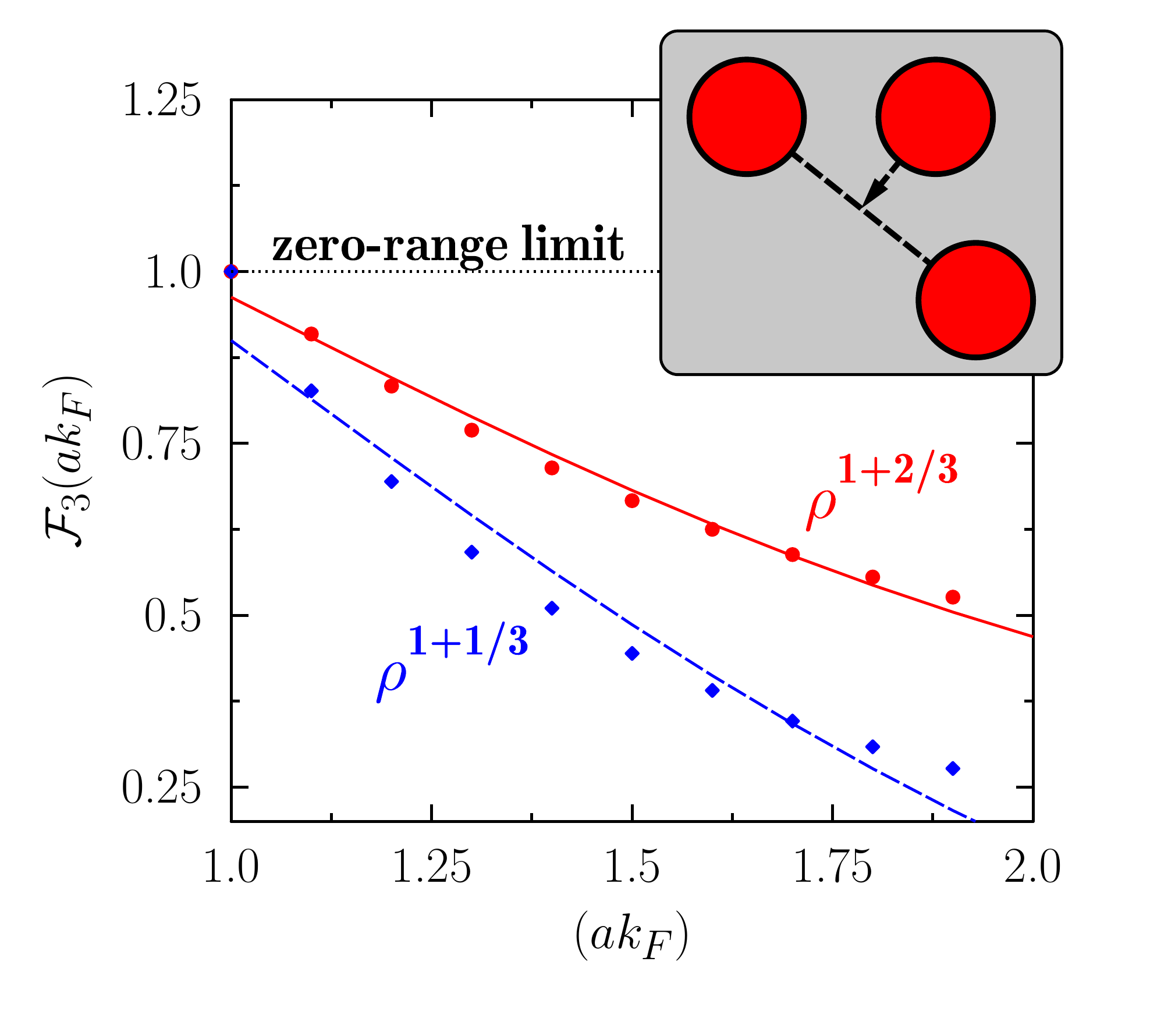}
\vspace{-5mm}
\caption{ (color online) ${\cal F}_3(k_Fa)$ as a function of $(ak_F)$ obtained
by adjusting the $c^{SM}_1$ and $c^{SM}_2$ to get a dependence similar to
$1/(a k_F)$ [$\alpha =2/3$] (red solid line) or $1/(a k_F)^2$ [$\alpha =1/3$]
(blue dashed line).  The reference curves $1/(a k_F)$ (red filled circles)
and $1/(a k_F)^2$ (blue filled diamonds) are also shown. The horizontal dotted
line indicates the zero-range 3-body interaction limit [$\alpha =1$].
The fit has been made for $(ak_F) \in [1.25,1.75]$, note that in nuclear
physics $(ak_F)\simeq 1.5$ at saturation.}
\label{fig:fitrho} 
\end{figure}
  
In the present work, we are interested in intermediate density region
(around the equilibrium configuration) where deviation from the limits
(\ref{eq:3zero}) is anticipated for finite-range interaction. To further progress, we assume that $g$ is a normalized gaussian function $g( r ) = e^{-(r^2/a^2)}/(a \sqrt{\pi})^3$.  In that case, 
using the same technique as in Ref.~\cite{Ben13},
integrals in Eq. (\ref{eq:edf}) become analytical functions of $x=(ak_F)$,
leading to
\begin{eqnarray}
\frac{{E}^{\rho\rho\rho}}{A} &=&  \frac{\rho^2}{6}  \left\{ c_1\left[ 1 - c_3F_1(x)
\right]+c_2  \left[  F_1(x/2) - c_3  F_2(x) \right] \right\} \nonumber \\
&\equiv& \frac{\rho^2}{6}\, {\cal F}_3(ak_F)\, , \label{eq:ef3}
\end{eqnarray}
with
\begin{align}
F_1 (x)=&\frac{12}{x^6}\left(1-e^{-x^2}\right)
-\frac{6}{x^4}\left(3-e^{-x^2}\right)
+\frac{6\sqrt{\pi}}{x^3}\,\Erf(x)\, , \nonumber
\end{align}
and
\begin{align}
F_2 (x)\!=&\frac{288}{x^8}\left(-e^{-\frac{x^2}{4}}-e^{-x^2}\right)
-\frac{24}{x^6}\left(12+e^{-x^2}-7e^{-\frac{x^2}{4}}\right) \nonumber \\
+& \frac{12}{x^4}\left(4e^{-\frac{x^2}{4}}-7e^{-x^2}\right)+\frac{6\sqrt{\pi}}{x^3}\left[8 \Erf(x)-7\Erf\left(\tfrac{x}{2}\right)\right] \nonumber \\
+&\frac{36\sqrt{\pi}}{x^7}\left(\frac{4}{x^2}-9\right)
\left[\Erf(x)-2\Erf\left(\tfrac{x}{2}\right)\right]\,. \nonumber 
\end{align}
Therefore, a density dependence $\rho^{1+\alpha}$ for the energy per particle
(as given by the commonly used density dependent Skyrme or Gogny interactions)
can be locally obtained if
${\cal F}_3(ak_F)  \propto (ak_F)^{3(\alpha - 1)} \propto \rho^{\alpha -1}$.
Adjusting the parameters $c_1$ and $c_2$, our semi-contact 3-body interaction can
thus approximate the desired density dependence in a given range
of the density. This is illustrated in Fig.~\ref{fig:fitrho}, for $\alpha =1/3$
and $\alpha=2/3$.  
The functional associated with the 3-body force can fairly well reproduce
the effect of a density dependent interaction with a non-integer power
of the density. Systematic analyses have shown that the 
present interaction is suitable for density dependence $\rho^{1+\alpha}$ of the energy per particle with $0 \leqslant \alpha \leqslant 1$. However, for very small $\alpha$, {\it i.e.} $\alpha \leqslant 1/6$, the fit starts to deteriorate.

\begin{figure}[htbp]
\vspace{-3mm}
\includegraphics[width=0.8\linewidth]{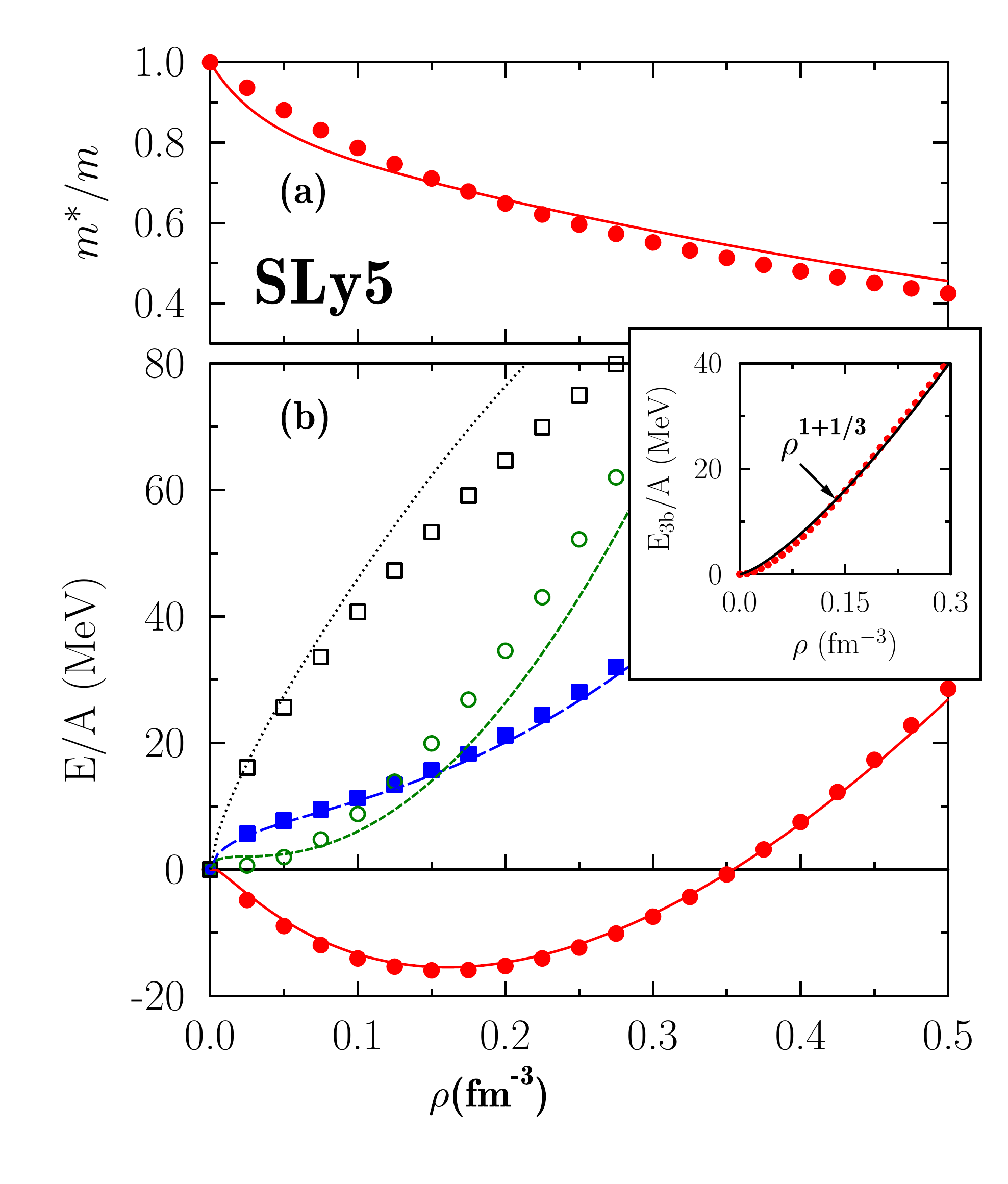}
\vspace{-6mm}
\caption{ (color online) Scalar-Isoscalar effective mass (a) and equations of
state (b) of SM (red), NM (blue), PM (green) and PNM (black) as a function
of density obtained with the SLy5 functional (markers) and with the Skyrme
[2-body] + 3-body functional (lines), called SLy5$^{\rm 3b}$. In the inset,
the contribution of the 3-body functional to the energy per particle in
symmetric matter is represented. A simple function proportional to $\rho^{1+\alpha}$ has been adjusted on it with $\alpha\simeq 1/3$ for an optimal fit
in the vicinity of saturation density.}
\label{fig:eossly5} 
\end{figure}

One of the recurrent difficulties of nuclear DFT based on zero-range
interactions is the impossibility to conjointly get the proper EOSs in infinite matter and a reasonable  behavior of the effective
mass~\cite{Boh79,Chab97}. The use of finite-range 3-body interaction
automatically induces a contribution to the effective mass given by
\begin{equation}
\frac{\hbar^2}{2m^*}=
  \frac{\rho^2_0a^2}{4} \left[
\frac{c_1^{\rm SM}}{24} M_1 \left(\frac{x}{2} \right)
  - \frac{c_2^{\rm SM} }{3}M_1 (x) + \frac{c_2^{\rm SM}}{8} M_2 (x) 
\right] \nonumber ,
\end{equation}
in SM, with
\begin{equation}
M_1 (x)= \frac{12}{x^6}\left(e^{-x^2}-1\right)
      + \frac{6}{x^4}\left(e^{-x^2}+1\right) ,\nonumber
\end{equation}
and
\begin{align}
M_2 (x)&= \frac{576}{x^8}\left(e^{-x^2}-e^{-\frac{x^2}{4}}\right)
       + \frac{288}{x^6} \nonumber \\
       &+ \frac{144\sqrt{\pi}}{x^7}\left(\frac{2}{x^2}-3\right)
         \left[2\Erf\left(\tfrac{x}{2}\right)-\Erf x\right]. \nonumber
\end{align}

As it can be done for the energy, the 3-body interaction parameters can be
adjusted to  reproduce specific behavior of the effective mass as a function
of density. For instance, it might be used to get the proper $\rho^{2/3}$
dependence  predicted by the Galitskii formula (see Eq. (11.62) of
Ref.~\cite{Fet03} and Ref.~\cite{Gez12} for a recent discussion).
An alternative situation is presented below, where we show that the 3-body
semi-contact interaction conjointly used with a 2-body density independent
interaction can appropriately describe the Skyrme prescription, {\it i.e.}
$(m^*/m) \propto \left( 1 + \theta \rho \right)^{-1}$ where the expression of
the parameter $\theta$ can be found in Ref.~\cite{Vau72}.  

In the nuclear context, it has recently become evident that the LDA-DFT 
with non-integer powers of the density leads to severe problems
to describe fluctuations along collective coordinates or restore broken
symmetries~\cite{Dug09}. Here, we will show that the semi-contact 3-body
interaction can replace the density dependent term used with standard
functionals.
To be considered as a practical tool, we should be able to find
a set of parameters that (i) provides a reasonable description of all
spin/isospin channels simultaneously (ii) conjointly describes the expected
density dependence of the effective mass. 
In the past, requirements (i) and (ii) have been successfully fulfilled
using a density dependent term. More recently, a zero-range density independent
interaction that could
meet these criteria was introduced~\cite{Sad13} using 3-body velocity
dependent terms. While possibly more difficult to implement, the semi-contact
3-body interaction limits the need of velocity dependent terms with the
advantage that it might give a better control on the unwanted finite-range
instabilities~\cite{Hel13}.

\begin{figure}[htbp]
\vspace{-3mm}
\includegraphics[width=0.8\linewidth]{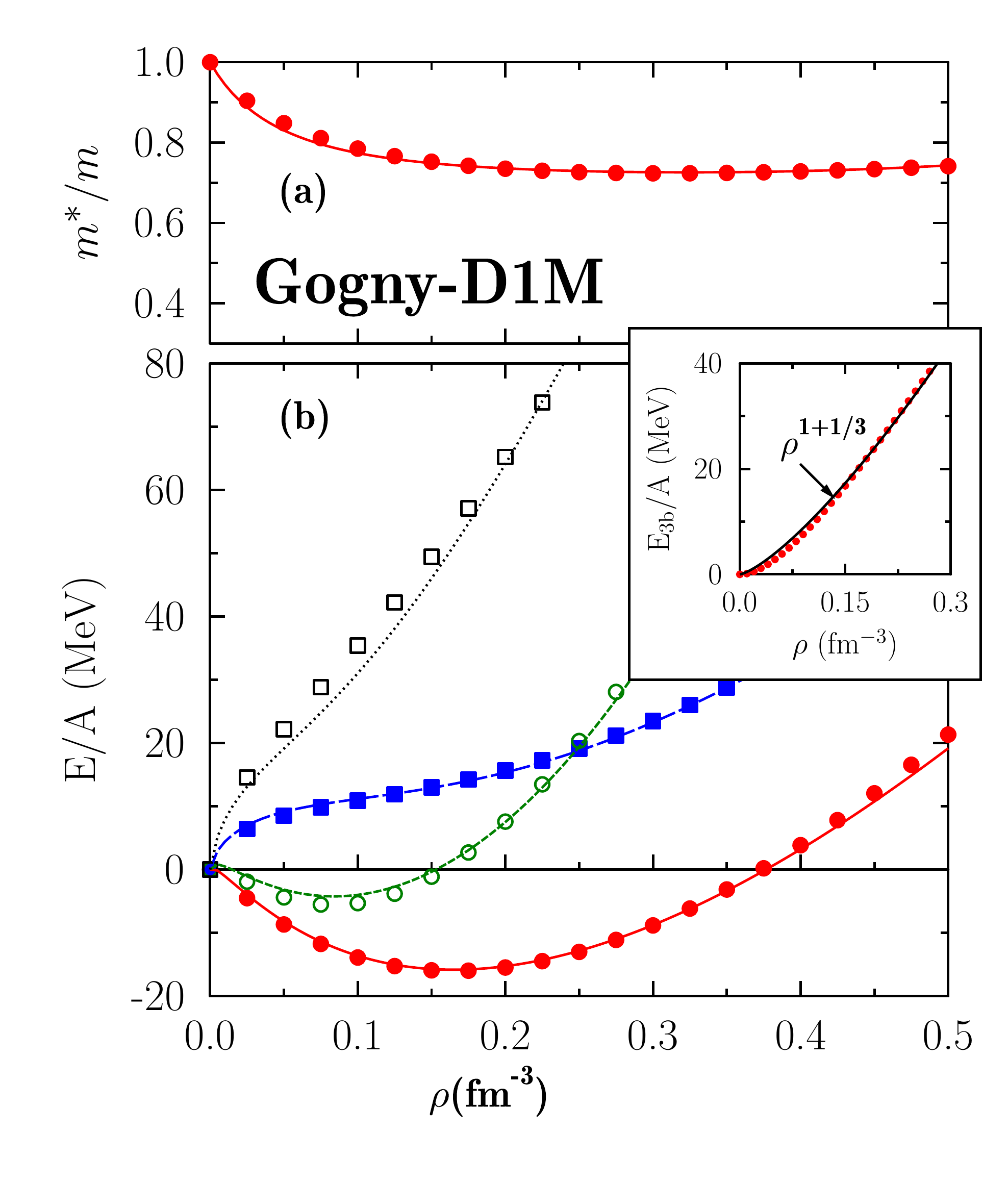}
\vspace{-6mm}
\caption{ (color online) Same as figure \ref{fig:eossly5} except that the
Gogny-D1M (markers) and D1M$^{\rm 3b}$ (lines) are shown. In the insert,
again $\alpha \simeq 1/3$ is found as optimal.}
\label{fig:eosgogny} 
\end{figure}

To illustrate that the new interaction can be employed successfully in the nuclear DFT context,  we have considered two commonly used 
functionals based on density dependent interaction, namely a Skyrme
(zero-range)~\cite{Sky56,Sky59,Vau72} and a Gogny
(finite-range)~\cite{Gog75,Gog75c} functional.  More precisely, we respectively 
considered the SLy5~\cite{Chab98} and D1M~\cite{Chap08} sets of parameters.  
These two functionals are known to provide meaningful description of the EOSs in different spin-isospin channels as well as reasonable effective mass
(see Figs.~\ref{fig:eossly5} and~\ref{fig:eosgogny}) and serves 
below as reference EOSs. 
\begin{table*}[htbp]
  \centering
\begin{tabular}{cccc}
\hline
     $t_0$(GeV fm$^3$) &    $t_1$ (GeV fm$^5$)&     $t_2$ (GeV fm$^5$) & \\
-1.283 (0.05\%)&0.874 (0.18\%)&-0.808 (0.69\%)& \\
\hline
     $x_0 $ &    $x_1$ &     $x_2$ & \\
0.29 (1.8\%)&0.51 (1.8\%)&-1.08 (0.03\%)& \\
\hline
     $v_0$  (GeV) & $v_\sigma$  (GeV)&  $v_\tau$  (GeV)& $v_{\sigma\tau}$  (GeV) \\
-15.03 (0.56\%)&24.13 (0.82\%)&24.99 (0.65\%)&-42.12 (0.25\%) \\
\hline
\end{tabular} 
\begin{tabular}{cccc}
\hline
     $W_1 $ (GeV) &    $B_1$ (GeV) &     $H_1$ (GeV) &     $M_1$ (GeV) \\
7.60 (1.3\%) & 2.54 (5.4\%) & 2.18 (5.8\%)  &  -1.47 (8.0\%) \\
\hline
     $W_2$ (MeV) &    $B_2$ (MeV) &     $H_2$ (MeV) &     $M_2$ (MeV) \\
-1047.97 (0.80\%) &  -86.54 (10\%) &  -681.06 (1.3\%) &     47.07 (10\%)  \\
\hline
     $v_0$  (GeV) & $v_\sigma$ (GeV) &  $v_\tau$ (GeV) & $v_{\sigma\tau}$ (GeV) \\
-14.65 (1.1\%) &  30.06 (0.82\%) & 19.02 (0.96\%) &  -42.71 (0.49\%) \\
\hline
\end{tabular}
  \caption{Parameters of the SLy5$^{\rm 3b}$ (left) and D1M$^{\rm 3b}$ (right). We use here standard  for the 2-body interaction parameters, see  Refs.~\cite{Chab98} and~\cite{Chap08}. In both cases, the range is fixed to $a=1.2$ fm. In the Gogny force case, the finite-ranges of the 2-body interaction have been kept equal to their original D1M values,  $\mu_1=0.5$~fm and $\mu_2=1.0$~fm. The uncertainty on parameters values, indicated in percent have been obtained 
  using standard covariance analysis. }
  \label{table:param}
\end{table*} 

We consider the original functionals and replace the density dependent term
by the functional deduced from the 3-body semi-contact interaction.
Doing so, the energy density functional can then be truly considered as the Hartree-Fock functional derived from 
a many-body Hamiltonian.
Then a global fit is made using the new functionals that are respectively labelled SLy5$^{\rm 3b}$ and D1M$^{\rm 3b}$ below. The parameters have been adjusted to reproduce the four EOSs (SM, NM, PM and PNM) and the (scalar-isoscalar)
effective mass simultaneously. The details of the fitting protocole are given
in the supplemental material~\cite{Sup14}.
A range $a=1.2$~fm, which is optimal for the fit on the D1M results, was chosen for the range of the semi-contact term.
Note that all parameters have been used for the fit except 
the ranges of the 2-body part that have been kept equal to the original D1M case. Altogether, the fit was made using respectively 10 and 12 parameters
for the   SLy5$^{\rm 3b}$ and D1M$^{\rm 3b}$. Optimal parameters values are given in table~\ref{table:param}. As it can be seen on both Figs.~\ref{fig:eossly5}
and \ref{fig:eosgogny}, an accurate reproduction of the original EOSs is
obtained along with the proper density dependence of 
the effective mass. In particular, the spin polarized matter properties that is difficult to grasp if a strict zero-range interaction is used, is well accounted for.
We see that the properties that characterize
the saturation point in SM, {\it i.e.} saturation density $\rho_\mathrm{sat}$,
binding energy per particle $B$, effective mass $m^*/m$ and compression
modulus $K_\infty$, are reproduced with deviations between targeted and obtained
values that are typical for nuclear EDFs. The values of these quantities 
obtained from SLy5 and D1M are compared with those computed from the new
functionals in table~\ref{tab:param}.

The new functionals SLy5$^{\rm 3b}$ and D1M$^{\rm 3b}$, where the density dependent term is replaced by the semi-contact 3-body interaction, provides a very good
reproduction of the saturation properties of the reference functionals, SLy5 and D1M, for all considered cases. In particular, the compression modulus
that was one of the motivations for the introduction of the $\rho^\alpha$ term
in the Skyrme and Gogny interactions has a reasonable value. 

\begin{table}[htbp]
  \centering 
\begin{tabular}{l|c |c ||c|c | }
\cline{2-5}
                               & ~~SLy5~~     & ~~SLy5$^{\rm 3b}$~~    &  ~~D1M ~~&  D1M$^{\rm 3b}$   \\
\hline
\multicolumn{1}{|l|}{$\rho_\mathrm{sat}$ [fm$^{-3}$]} &  0.160 & 0.161&  0.165 &    0.165    \\
\hline
\multicolumn{1}{|l|}{$B/A$ [MeV] }  & -15.98           & -15.42 & -16.02 &  -15.82 \\
\hline
\multicolumn{1}{|l|}{$K_\infty$ [MeV] }  &  229.92 & 236.59 &      224.98  &  228.58  \\
\hline
\multicolumn{1}{|l|}{$m^*/m$}  & 0.697        & 0.691 & 0.746& 0.744 \\
\hline
\end{tabular}
  \caption{Values of the saturation density $\rho_\mathrm{sat}$, binding energy $E/A$, compression modulus 
 $K_\infty$  and isoscalar effective mass $m^*/m$ for the functionals considered in this work.}
  \label{tab:param}
\end{table}

In the present work, a simplified 3-body interaction is used to  construct a density functional theory. It is shown that
this new interaction can mimic the behavior of density dependent interactions
used in the nuclear 
context. This interaction solves two important difficulties encountered in the applicability of nuclear DFT. First, it provides a correct description of 
saturation properties and 
a reasonable  description of the infinite matter equations of state
in various spin-isopin channels  as well as the effective masse density dependence for a wide range of densities.
Second, by replacing the pathological term $\rho^\alpha$ in the functional by a
density independent interaction,  the new functionals can be used in a DFT Multi-Reference framework for instance to restore symmetries or to account of configuration mixing beyond the 
independent particle picture.

\begin{acknowledgments}
We thank M. Bender, J. Dobaczewski, T. Duguet and M. Kortelainen for
discussions. This work has been supported in part by the Academy of
Finland and University of Jyv\"askyl\"a within the FIDIPRO programme.
\end{acknowledgments}


\begin{thebibliography}{99}


\bibitem{Lee57} T. D. Lee and C. N. Yang, Phys. Rev. {\bf 105}, 1119 (1957). 

\bibitem{Fet03} A.~L. Fetter and J.~D. Walecka, {\it Quantum Theory of Many-Particle Systems} (McGraw-Hill, New York, 1971). 

\bibitem{Bul07} A. Bulgac, Phys. Rev. {\bf A 76}, 040502R (2007).

\bibitem{Gez08} A. Gezerlis and J. Carlson, Phys. Rev. C {\bf 77}, 032801(R) (2008).

\bibitem{Gez10} A. Gezerlis and J. Carlson, Phys. Rev. C {\bf 81}, 025803(R) (2010).

\bibitem{Ben03} M.~Bender, P.-H.~Heenen, and P.-G.~Reinhard, Rev. Mod. Phys. {\bf
75}, 121 (2003).

\bibitem{Rin80} P. Ring and P. Schuck, {\it The Nuclear Many-Body Problem}
(Springer-Verlag, Berlin, 1980).

\bibitem{Bla86} J.P. Blaizot and G. Ripka, {\it Quantum Theory of Finite Systems}, (MIT Press, Cambridge, Massachusetts, 1986).

\bibitem{Lac09} D. Lacroix, T. Duguet, and M. Bender, Phys. Rev. {\bf C 79}, 044318
(2009).

\bibitem{Ben09} M. Bender, T. Duguet, and D. Lacroix, Phys. Rev. {\bf C 79}, 044319
(2009).

\bibitem{Dob07d} {J. Dobaczewski, M.V. Stoitsov, W. Nazarewicz, and P.-G. Reinhard, Phys. Rev. C
  {\bf 76}, 054315 (2007)}.

\bibitem{Dug09} T. Duguet, M. Bender, K. Bennaceur, D. Lacroix, and
T. Lesinski, Phys. Rev. {\bf C 79}, 044320 (2009).

\bibitem{Was12}
{K. Washiyama, K. Bennaceur, B. Avez, M. Bender, P.-H. Heenen, V. Hellemans,
  Phys. Rev. C {\bf 86}, 054309 (2012)}.
  
\bibitem{Sad13} {J. Sadoudi, T. Duguet, J. Meyer, and M. Bender, Phys. Rev. C {\bf 88}, 064326
  (2013)}.


\bibitem{Dob12a}
{J. Dobaczewski, K. Bennaceur, and F. Raimondi, J. Phys. G {\bf 39}, 125103 (2012)}.

\bibitem{Ben13}
{K. Bennaceur, J. Dobaczewski, and F. Raimondi,
EPJ Web of Conferences {\bf 66}, 02031 (2014)}.

\bibitem{Rai14}
{F. Raimondi, K. Bennaceur, and J. Dobaczewsk, J. Phys. G {\bf 41}, 055112 (2014)}.

\bibitem{Gez10b}{A. Gezerlis, and G. F. Bertsch, Phys. Rev. Lett. {\bf 105}, 212501 (2010)}.

\bibitem{Boh79} O. Bohigas, A. M. Lane, and J. Martorell,
Phys. Rep. {\bf 51}, 267 (1979).

\bibitem{Chab97} E. Chabanat, P. Bonche, P. Haensel, J. Meyer, and R. Schaeffer, Nucl. Phys. {\bf A627}, 720 (1997).


\bibitem{Gez12} A. Gezerlis and G. F. Bertsch, Phys. Rev. {\bf C85}, 037303 (2013).

\bibitem{Vau72} D. Vautherin and D. M. Brink, Phys. Rev. C 5, 626 (1972).


\bibitem{Hel13} V. Hellemans, A. Pastore, T. Duguet, K. Bennaceur, D. Davesne, J. Meyer, M. Bender, and P.-H. Heenen,
Phys. Rev. {\bf C 88}, 064323 (2013).

\bibitem{Sky56}
{T.H.R. Skyrme, Phil. Mag. {\bf 1}, 1043 (1956)}.

\bibitem{Sky59}
{T.H.R. Skyrme, Nucl. Phys. {\bf 9}, 615 (1959)}.

\bibitem{Gog75}
{D. Gogny, Nucl. Phys. {\bf A237}, 399 (1975)}.

\bibitem{Gog75c}
{D. Gogny, Proceedings of the International Conference on Nuclear
  Selfconsistent Fields, Trieste 1975. G. Ripka and M. Porneuf. Eds North
  Holland. Amsterdam, 176, 209, 265, 266 (1975)}.

\bibitem{Chab98} E. Chabanat, P. Bonche, P. Haensel, J. Meyer, and R. Schaeffer, Nucl. Phys. {\bf A635}, 231 (1998).

\bibitem{Chap08} F. Chappert, M. Girod, and S. Hilaire, Phys. Lett. {\bf B668},
420 (2008).


\bibitem{Sup14} See Supplemental Material for the details of the fitting
protocole.


\end{thebibliography}
\end{document}